\documentclass{ws-procs9x6}

\usepackage{bm}
\DeclareBoldMathCommand{\bfxi}{\xi}
\DeclareBoldMathCommand{\bfpi}{\pi}

\begin{document}

\title{On the Casimir entropy between `perfect crystals'}

\author{C. HENKEL$^*$ and F. INTRAVAIA\footnote[4]{Present 
address: Theoretical Division, MS B213,
Los Alamos National Laboratory, 
Los Alamos NM 87545, U.\ S.\ A.}}

\address{Institut f\"ur Physik und Astronomie,
Universit\"at Potsdam,\\ 
Karl-Liebknecht-Str. 24/25,
14476 Potsdam, Germany\\
$^*$E-mail: henkel@uni-potsdam.de
}

\begin{abstract}
We give a re-interpretation of an `entropy defect' in the electromagnetic
Casimir effect. The electron gas in a perfect crystal
is an electromagnetically disordered system
whose entropy contains a finite Casimir-like contribution. The Nernst theorem
(third law of thermodynamics) is not applicable.
\end{abstract}


\bodymatter

\section{Introduction}
	\label{s:intro}

%









It is well known that fluctuation interactions at nonzero temperature are 
entropic in character, a prominent example being the critical Casimir 
effect in liquid mixtures close to a continuous phase transition (see Ref.\refcite{Gambassi09}
for an overview). The electromagnetic Casimir interaction is also associated 
with an entropy that determines its limiting behaviour at high
temperatures and/or large distances\cite{Balian78, Feinberg01}. 
The Casimir entropy for two material plates has recently attracted much
interest also for low temperatures, as for certain situations a violation of the
third law of thermodynamics (the Nernst heat theorem) has been 
claimed\cite{Klimchitskaya01a, Bezerra04}. 
This has been used to argue in favor of a description 
where the DC conductivity of the metallic plates
is ignored. Although the result of this theoretical prescription provides
a better 
fit to recent experiments\cite{Decca07b}, the situation is, however, not satisfactory from the physical point of view.
In addition, a similar analysis for an experiment with laser-irradiated 
semiconductors\cite{Chen07a} leaves open the meaning of the threshold
value above which the DC conductivity should be included in the theory.

Much has been said about spatially dispersive mirrors where the third law is
verified, due to the anomalous skin effect\cite{Svetovoy08b}, and where
a continuous cross-over from a dielectric to a perfectly conducting response 
has been found\cite{Pitaevskii08}. We focus in these proceedings on a
strictly local framework, mainly for simplicity, but also to show that this 
case is thermodynamically consistent as well.
We shall see that, indeed, spatial
dispersion plays only a small role in the range of wave vectors that are relevant 
for current Casimir experiments.

We take up the interpretation of Ref.\refcite{Hoye03a} where the nonzero
Casimir entropy found as $T \to 0$ was associated to two oscillators coupled 
via a third one. Following this idea, we consider two half-spaces filled by an
ideal electron gas, separated by a distance $L$,
and provide a direct calculation of the entropy per area
$S(L) = \lim\limits_{T\to 0} S(L, T)$ in one of the two field polarizations.
This calculation highlights the following point: 

The Casimir entropy $S(L)<0$ results from the coupling between two 
systems that are not in equilibrium as $T \to 0$. They are filled with a \emph{frozen}
magnetization and, in the local limit, have separately divergent (bulk and
surface) entropies that characterize the disorder of the electromagnetic
configuration.
The Casimir entropy is the correction to additivity when the two bodies are close enough 
for their frozen currents to be mutually
coupled by the quasi-static magnetic fields that
`leak' through their surfaces. The Nernst theorem is clearly not applicable for 
this disordered system. The situation is quite similar to the `ideal conductor' 
(in distinction to a superconductor, see Ref.\refcite{Landau10}) that does not
reach thermodynamical equilibrium as it is cooled, because its random
bulk currents freeze. (See Ref.\refcite{Evans65} where it is argued how
special this ideal conductor case is.)

\section{Casimir entropy from frozen medium currents}


\subsection{Motivation}

We have analyzed in a recent paper\cite{Intravaia09a} the overdamped
field modes to which the unusually large thermal corrections to the 
Casimir force between metals can be attributed. Substantiating
previous observations\cite{Torgerson04,Svetovoy07}, we have
interpreted the characteristic frequency, 
$\xi_L = D / L^2 = (\mu_0 \sigma L^2)^{-1}$, 
in terms of a diffusion equation (diffusion coefficient $D$)
satisfied by the magnetic field and electric
currents in a medium with DC conductivity $\sigma$. 

There is no contradiction between thermodynamics and fluctuation 
electrodynamics in this case. Fields and
currents induced in the metal are clearly damped and lose energy into
the phonon bath, say. 
In equilibrium, however, this is compensated by field and current fluctuations
that are created by the bath. This concept can be traced back to the
Einstein--Langevin theory of Brownian motion \cite{Einstein05b} and
is also the very essence of the fluctuation-dissipation 
theorem\cite{Callen51,Ford93a}.  Moreover, for the quantum field
theory, the quantum (or zero-point) fluctuations of the bath variables
are an essential tool to establish at all times the commutation relations for 
the field operators\cite{Barnett92,Welsch95,Tip97,Suttorp07,Scheel08}.
In the field theory considered in Ref.\refcite{Intravaia09a}, we dealt with 
overdamped 
modes: if the wave equation were homogeneous, its eigenfrequencies
would be purely imaginary, similar to free Brownian particles. The quantum
theory of Brownian motion\cite{WeissBook} provides a consistent scheme
for the quantum thermodynamics of this damped system.
In this setting, nonzero entropies and even negative heat
capacities find a quite natural explanation (see, e.g.,
Refs.\refcite{Ingold09a,Ingold09b}).

In the particular
case of an ideal electron gas (or `perfect crystal'), the diffusion constant 
$D = D(T) =
\mathcal{O}(T^2)$. As the temperature drops to zero,
the diffusion-dominated modes of
the electromagnetic field do not reach a unique ground state, but remain
in the classical regime $\hbar\xi_L \ll T$. 
This motivates the present calculation where the electromagnetic Casimir 
entropy between
perfect crystals is re-derived within a classical model.

%
The basic ingredient are
transverse modes that extend throughout the bulk of the gas:
static current waves interlocked with a magnetic field. 
The magnetic fields associated to the bulk currents
leave one medium, by continuity, and cross the vacuum gap to the
other medium in the form of (transverse) evanescent waves. This coupling
between the two media changes slightly the wave vector of each current
mode. Summing over all modes, we get a non-zero change in entropy
that depends, quite naturally, on the separation $L$. It represents
the distance-dependent change per unit area of the (much larger) entropy
of the two frozen bulk systems. 

\subsection{Lagrangian and conservation laws}

We start with the Lagrangian density
\begin{equation}
	\mathcal{L} = 
	\frac{ n m }{ 2  } 
    \dot {\bfxi}^2 
    + e n \dot {\bfxi} \cdot {\bf A} 
    - \frac{1}{2\mu_{0}}( \nabla \times {\bf A})^2
	\label{ch:def-class-Lagrangian}
\end{equation}
where the field $\bfxi$ describes the displacement of a charged fluid
element, ${\bf A}$ is the vector potential, 
$n$ a constant background charge density and $e$ a coupling constant
with units of charge.  
The current density is
${\bf j} = e n \dot{\bfxi}$ [see Eq.(\ref{ch:Maxwell-Faraday}) below], 
so that $\dot{\bfxi}$ represents a velocity
field. The first term in Eq.(\ref{ch:def-class-Lagrangian}) is thus the kinetic
energy (density), the second one a bilinear coupling, and the third one the 
magnetic energy. Note that we neglect electric fields here. This is consistent
if we make the assumption that
$\nabla \cdot \bfxi = \nabla \cdot {\bf A} = 0$. The first equality
ensures that the medium displacement does not produce any charge density,
the second one is the Coulomb gauge. 
The variation of
the Lagrangian~(\ref{ch:def-class-Lagrangian}) with respect to ${\bf A}$ gives 
the Faraday equation
\begin{equation}
	    e n \dot{\bfxi} 
    -\frac{1}{\mu_{0}} \nabla \times (\nabla \times {\bf A}) = 0
    \label{ch:Maxwell-Faraday}
\end{equation}
In addition, the displacement
field $\bfxi$ is a cyclic variable, hence we get a conserved momentum
field
\begin{equation}
	\frac{ \partial {\bfpi}}{ \partial t } = 0
, \quad
	{\bfpi} = 
	n m \dot \bfxi + e n {\bf A}
	\label{ch:conservation-law}
\end{equation}
There are two ways to implement this conservation law
physically.\cite{London35}


(i) In a \emph{London 
superconductor}, the current density is tied at all times to the vector potential,
with the momentum ${\bfpi}$ being zero:
\begin{equation}
	\mbox{superconductor:\ }\quad
	{\bf j} = e n \dot{\bfxi} =
	- \frac{n e^2}{m} {\bf A}
	\label{ch:London-equation}
\end{equation}
The Maxwell--Faraday equation~(\ref{ch:Maxwell-Faraday}) becomes
\begin{equation}
	\left( \lambda^{-2} -
	\nabla^2 \right) {\bf A} = 0
	\label{ch:Meissner-effect}
\end{equation}
where the Mei\ss{}ner--London penetration depth $\lambda$
is given by the
familiar expression
$\lambda^{-2} = \mu_0 n e^2 / m = \Omega^2 / c^2$ 
($\Omega$ is the plasma frequency). Eq.(\ref{ch:Meissner-effect}) does
only allow for solutions that start at the surface and exponentially 
decay into the bulk on a length scale $\lambda$ (or shorter). Except
for a surface layer of thickness $\sim \lambda$, the
interior of the medium remains free of magnetic field:
the Mei\ss{}ner--Ochsenfeld effect.

From the London 
equation~(\ref{ch:London-equation}), we can also conclude that
the Mei\ss{}ner effect is maintained in
time-dependent fields
(at least with sufficiently slow variations;
a detailed analysis clearly goes beyond the simple model considered here). 
 For a given frequency component $\omega$,
the `dielectric function' of the London superconductor can be read off
from the polarization field associated to $\bfxi$:
\begin{equation}
	{\bf P} = e n \bfxi = \frac{ {\bf j} }{ - {\rm i} \omega }
	= - \frac{ \varepsilon_0 \Omega^2 }{ \omega^2 }
	{\bf E}
	\label{ch:return-to-plasma-model}
\end{equation}
leading to the so-called plasma model
$\varepsilon( \omega ) = 1 - \Omega^2 / \omega^2$. There is no
violation of causality here, if we read Eq.(\ref{ch:London-equation}) 
as a retarded response function between the current density and the time 
integral of the electric field (i.e., the vector potential).


The option (ii) that complies with the conservation law~(\ref{ch:conservation-law})
corresponds to an \emph{ideal conductor}:
\begin{equation}
	\mbox{ideal conductor:\ }\quad
	\frac{ \partial {\bf j} }{ \partial t }  = 0 
	\quad \mbox{and} \quad
	\frac{ \partial {\bf A} }{ \partial t } = 0 
	\label{ch:ideal-conductor}
\end{equation}
which means that currents, once created, are not damped and that
the magnetic field is static. The value of the conserved momentum ${\bfpi}$ 
is not restricted otherwise.
The Faraday
equation~(\ref{ch:Maxwell-Faraday}) then yields the field ${\bf A}$ 
in terms of its source ${\bf j}$.
Note that the magnetic field is in this case tied to the current density,
similar to the scalar potential and the charge density
in Coulomb-gauge electrodynamics~\cite{CDG1}.
Let us switch to reciprocal space with wavevector ${\bf q}$:
the vector potential ${\bf A}_{\bf q}$ 
created by the current is
\begin{equation}
	{\bf A}_{\bf q} 
	=
	\mu_0 \frac{ {\bf j}_{\bf q} }{  {\bf q}^2 } 	
	\label{eq:A-in-terms-of-momentum}
\end{equation}
so that the Lagrangian~(\ref{ch:def-class-Lagrangian}) becomes
\begin{equation}
	L = \frac{\mu_0 V }{2} 
\sum_{\bf q}
\left(
\lambda^2
+ {\bf q}^{-2}
\right)
|{\bf j}_{\bf q}|^2
	\label{eq:L-in-terms-of-current-alone}
\end{equation}
where $V$ is the quantization volume.
The conjugate momentum becomes
\begin{equation}
	\bfpi_{\bf q} 
	= \frac{ m }{ e } \left( 1 + \frac{ 1 }{ \lambda^2 {\bf q}^2 }
	\right)
	{\bf j}_{\bf q}
	\label{eq:conjugate-momentum-and-current}
\end{equation}

\subsection{Normal modes and entropy} 

The normal modes of the effective 
Lagrangian~(\ref{eq:L-in-terms-of-current-alone}) can clearly be chosen
as plane waves, labelled by wave vector ${\bf q}$ and polarization
index $\mu$. 
The associated Hamiltonian, expressed
in terms of the canonical momentum field, is then
\begin{equation}
    H = 
    \frac{ V }{ 2 n m }
    \sum_{{\bf q},\mu}
    \left( 1 + 
    \frac{ 1 }{ \lambda^2 {\bf q}^2 } \right)^{-1} |{\pi}_{{\bf q}\mu}|^2
	\label{eq:glass-Hamiltonian}
\end{equation}
The (classical) thermodynamics of this system is determined by summing
the free energies of the normal modes over the
quantum numbers ${\bf q}$, $\mu$.
For one mode, we find by calculating the classical partition 
function
($\beta = 1/T$)
\begin{equation}
    F_{{\bf q}\mu} = - T \log 
    \int\!{\rm d}{\pi}_{{\bf q}\mu}^{\phantom *} 
    \exp\left( - \frac\beta2  \epsilon_{q} |\pi_{{\bf q}\mu}|^2 \right)
    = \frac{ T }{ 2 } \log\frac{ \beta \epsilon_{q} }{ 2 \pi }
    \label{eq:result-class-free-energy}
\end{equation}
where $\epsilon_q = 
(V/nm) (1 + \lambda^{-2}{\bf q}^{-2})^{-1}$ determines
the mode's energy.
The entropy of this polarization mode is
\begin{equation}
    S_{{\bf q}\mu} = - \frac{ \partial F_{{\bf q}\mu} }{ \partial T }
    = - \frac{ 1 }{ 2 } \log\frac{ \beta \epsilon_{q} }{ 2 \pi }
    + \frac{ 1 }{ 2 }
    \label{eq:entropy-per-mode-1}
\end{equation}
where the equipartition term $+1/2$ comes from the $\beta$ in the logarithm.

When we sum this over all modes, the entropy becomes to leading order 
extensive in the volume $V$ of the medium. The part that depends on
the surface is calculated in the usual way. Consider two media (total volume 
$V$) with parallel
surfaces of area $\mathcal{A}$ facing each other at a distance $L$ and write
the entropy of the total system in the form
\begin{equation}
    S = V s + 2 {\cal A} S_{\rm surf} + {\cal A} S(L)
        \label{eq:def-Casimir-entropy}
\end{equation}
where $s$ is the (intensive) bulk entropy density, $S_{\rm surf}$ 
is the entropy per area of one (isolated) surface and $S(L)$
the Casimir entropy per area. We can read the latter as the
deviation from additivity in the system of two media: it thus describes
how the disorder (or information content) of the two plates
is changed by the coupling across the vacuum gap of thickness $L$.

\begin{figure}[htb]
\centerline{\psfig{file=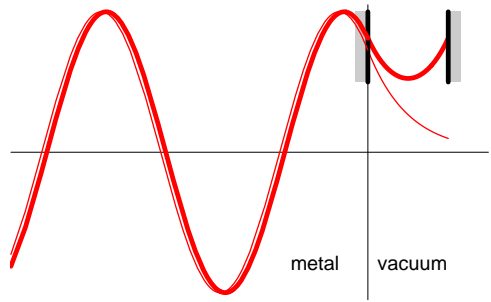,width=2.5in}}
\caption{Illustration of standing waves at the surface 
of an ideally conducting medium. We plot the component of the
vector potential tangential to the surface.
Thin line: isolated surface, thick line: 
mode between two surfaces with even parity.}
\label{ch:standing-waves}
\end{figure}

The physical mechanism for this coupling is the penetration of magnetic
fields through the medium surface, as allowed for by the
electromagnetic boundary conditions.  In the vacuum between the media, 
the fields satisfy
the Laplace equation $\nabla^2 {\bf A} = 0$: for a given wave vector
${\bf k}$ parallel to the surface, they `propagate' perpendicular to the
surface (along the $z$-axis, say) as evanescent waves $\sim
\exp( \pm k z )$ where $k = |{\bf k}|$.
For a single surface, only solutions that decay into
the vacuum are permitted. In the gap $0 \le z \le L$ between two surfaces,
even and odd solutions
$\cosh k (z - L/2 )$ and $\sinh k (z - L/2 )$ can be constructed.
This is illustrated schematically in Fig.\ref{ch:standing-waves}.

Both the surface entropy and the Casimir entropy can be calculated from
the phase shifts of standing wave modes 
(see Refs.\refcite{Barton79,Bordag01} for 
details). For the surface entropy,
\begin{equation}
    S_{\rm surf} = \sum_{\mu}
    \int\frac{ {\rm d}^2k }{ (2\pi)^2 }
    \int\limits_{0}^{\infty}\!{\rm d}k_{z}
    S_{{\bf q}\mu}
    \left( - \frac{ 1 }{ \pi }
    \frac{ \partial \theta_\mu }{ \partial k_{z} } \right)
    \label{eq:def-surface-entropy}
\end{equation}
where ${\bf q} = ({\bf k}, k_z)$, and
the mode functions in the medium ($z \le 0$)
are proportional to ${\rm e}^{ {\rm i} {\bf k} \cdot {\bf r}_\Vert }
\sin( k_z z + \theta_\mu )$ with ${\bf r}_\Vert = (x, y)$ the coordinates
parallel to the surface.
From this form, we can also read off 
a `reflection coefficient' for (time-independent) waves from within the 
medium, $r_\mu = - {\rm e}^{ - 2 {\rm i} \theta_\mu }$. 
From a physical point of
view, we can interpret the phase derivative in Eq.(\ref{eq:def-surface-entropy})
as a density of modes in ${\bf q}$-space, more precisely, its change due 
to the surface. 

For an isolated interface, the usual matching of the component
of the vector potential tangential to the surface and its derivative at the 
interface yields in TE-polarization (current perpendicular to the
plane of incidence spanned by ${\bf k}$ and the surface normal)
\begin{equation}
 	r_{\rm TE} = \frac{ k_z - {\rm i} k }{ k_z + {\rm i} k }, \qquad
	\tan \theta_{\rm TE} = - \frac{ k_z }{ k }
	\label{ch:TE-inside-reflection}
\end{equation}
Here, $k = |{\bf k}|$ gives the decay constant of the evanescent wave on
the vacuum side. In the TM-polarization (current in the plane of incidence),
the current has to satisfy the boundary condition $\lim\limits_{z \to 0}
j_z( z ) = 0$ to avoid the build-up of a surface charge sheet. (That case
would
require electrical field energy in the Lagrangian~(\ref{ch:def-class-Lagrangian})
and is best described within a spatially dispersive model.\cite{Barton79})
This boundary condition immediately leads to $r_{\rm TM} = 1$, and there
is no phase shift.
%
The surface entropy in TM-polarization hence vanishes, while it is
logarithmically divergent at large  ${\bf q}$
in the TE-polarization:
$S_{\rm surf} \approx -(8\pi\lambda^2)^{-1} \log( q_c \lambda )$
with a short-range cutoff $q_c$.
One needs a nonlocal description of the material response (spatial dispersion)
to get a finite result, see, e.g., Ref.\refcite{Horing85} for the surface self-energy.

For the Casimir entropy, a local calculation is sufficient, as we shall see now:
the reflection phases for even and odd modes in the vacuum gap
are found as (we henceforward suppress the TE-polarization label)
\begin{equation}
	\tan \theta_{\rm even}( L ) = - \frac{ k_z }{ k } \coth \frac{ k L }{ 2 },
	\qquad
	\tan \theta_{\rm odd}( L ) = - \frac{ k_z }{ k } \tanh \frac{ k L }{ 2 }
	\label{eq:even-odd-phase}
\end{equation}
The entropy per area for the two-surface system,
$2 S_{\rm surf} + S(L)$,
 is then given by
Eq.(\ref{eq:def-surface-entropy}) with $\theta$ replaced by
$\theta_{\rm even}( L ) + \theta_{\rm odd}( L )$. Subtract
twice the single-interface phase shift and calculate the quantity
\begin{equation}
	\exp 2 {\rm i}[ \theta_{\rm even}( L ) + \theta_{\rm odd}( L ) - 
	2 \theta] = 
	\frac{ 1 - r^2 \, {\rm e}^{ - 2 k L }
	}{ 1 - (r^*)^2 \, {\rm e}^{ - 2 k L }  }
	\label{eq:get-Casimir-phase-shift}
\end{equation}
as can be checked with straightforward algebra. 
The Casimir entropy from
TE-polarized bulk currents becomes 
[combining Eqs.(\ref{eq:entropy-per-mode-1},
\ref{eq:def-surface-entropy},
\ref{eq:get-Casimir-phase-shift})]
\begin{equation}
	S(L) = 
    \int\limits_{0}^{\infty}\!\frac{ k {\rm d}k }{ 2\pi }
    \int\limits_{0}^{\infty}\!\frac{ {\rm d}k_{z} }{ 2\pi }
\left( 
\log\frac{ \beta V \lambda^2 {\bf q}^2 }{ 2 \pi n m 
(1 + \lambda^{2}{\bf q}^{2})} 
-1
\right)
    \frac{ \partial  }{ \partial k_{z} } 
    {\rm Im} \, \log
    \left( 
1 - r^2 \, {\rm e}^{ - 2 k L }
    \right)
	\label{eq:Casimir-entropy-2}
\end{equation}
The terms independent of $k_z$
in the entropy per mode are irrelevant:
after a partial integration, the integrated terms vanish because 
for $k_z \to 0, \infty$, the reflection coefficient $r$ becomes real.
Manifestly, short-wavelength modes with $2 k L \gg 1$ are suppressed,
and a local theory is sufficient unless $L$ becomes comparable to the 
length scales typical for spatial dispersion (mean free path, Debye-H\"uckel
screening length, Fermi wavelength).

\subsection{Calculation of the entropy}

\begin{figure}[htb]
\centerline{\psfig{file=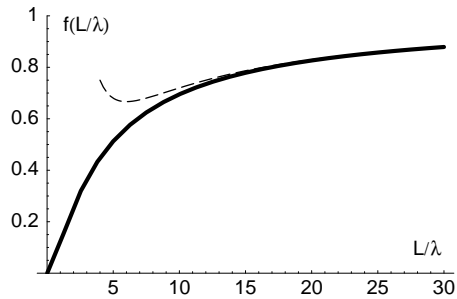,width=2.5in}}
\caption{Scaling function $f(L/\lambda)$
for the Casimir entropy of for the ideal electron gas, defined by 
Eqs.(\ref{eq:result-Foucault-entropy}, \ref{eq:def-scaling-function}).
Dashed: Eq.(\ref{eq:series-scaling-function}). $\lambda = c / \Omega$ 
is the London-Mei\ss{}ner penetration depth (plasma wavelength).
The same $f(L/\lambda)$ governs the `entropy defect' for the 
electromagnetic Casimir effect between perfect crystals
(i.e., Lifshitz theory with the Drude dielectric function
and scattering rate $\gamma(T) = \mathcal{O}(T^2)$),
see Eq.(20) in Ref.\refcite{Bezerra04}.}
\label{ch-fig:scale-function}
\end{figure}

Integrating Eq.(\ref{eq:Casimir-entropy-2}) 
by parts, we have to evaluate the integral:
\begin{eqnarray}
I_k &=& -
	\int\limits_{-\infty}^{\infty}\!\frac{ {\rm d}k_{z} }{ 2\pi {\rm i} 
	\,\lambda^2 }
    \frac{ k_z \,
	    \log
	    \left( 1 - r^2 \, {\rm e}^{ - 2 k L } \right)
    }{ (k_z^2 + k^2) (k_z^2 + k^2 + \lambda^{-2}) }
\end{eqnarray}
where we recall that $r$ is given by Eq.(\ref{ch:TE-inside-reflection}) above.
We have extended the integration domain to $-\infty < k_z < +\infty$,
using the property $r(-k_z) = [r(k_z)]^*$. Observe that $r(k_z)$,
as a function of complex $k_z$, satisfies $|r(k_z)| \le 1$ 
in the upper half-plane,
that the integrand vanishes at infinity, and evaluate the integral by
closing the contour. There are simple poles at $k_z = {\rm i} k$
and $k_z = {\rm i} (k^2 + \lambda^{-2})^{1/2}$. At the first pole,
the reflection coefficient~(\ref{ch:TE-inside-reflection}) 
vanishes, and we get from the second one:
\begin{equation}
	I_k = \frac{ 1 }{ 2 }
	\log\left[ 1 - r_{\rm pl}^2( k, 0 ) {\rm e}^{ - 2 k L } \right]
	,\quad
	r_{\rm pl}( k, 0 ) = 
	\frac{(k^2 + \lambda^{-2})^{1/2} - k }{
	(k^2 + \lambda^{-2})^{1/2} + k}
	\label{eq:result-integral-I-k}
\end{equation}
As it happens, the reflection coefficient $r_{\rm pl}(k, \omega )$ 
for electromagnetic waves from a plasma half-space
[dielectric function after Eq.(\ref{ch:return-to-plasma-model})] 
appears here, evaluated at zero frequency and in the TE-polarization. 
If Eq.(\ref{eq:result-integral-I-k}) is integrated over $k$, we get 
the `entropy defect' calculated in the Lifshitz theory of the Casimir effect
using the local dielectric function of a `perfect crystal' [see, e.g.,
Eq.(20) of Ref.\refcite{Bezerra04}]:
\begin{equation}
	S(L) = \int\limits_{0}^{\infty}\!\frac{ k {\rm d}k }{ 2\pi } I_k
	=
	    \int\limits_{0}^{\infty}\!\frac{ k {\rm d}k }{ 4\pi }
	\log\left[ 1 - r_{\rm pl}^2( k, 0 )\,
	{\rm e}^{ -2 k L } \right]
	\label{eq:result-Foucault-entropy}
\end{equation}
A switch to the integration variable to $y = 2 k L$ shows that 
\begin{equation}
	S(L) = -\frac{ \zeta(3) }{  16\pi } \frac{ f(L/\lambda) }{ L^2 }
	\label{eq:def-scaling-function}
\end{equation}
where the scaling function $f(L/\lambda)$, plotted in 
Fig.\ref{ch-fig:scale-function},
is dimensionless and
normalized to unity in the limit $L \gg \lambda$. Indeed, in this regime,
one may expand $r_{\rm pl}(k, 0 )$ in powers of $k$ to get the asymptotic
series [for higher terms, see Eq.(20) of Ref.\refcite{Bezerra04}]:
\begin{equation}
	L \gg \lambda
	: \quad
	f( L / \lambda) \approx 1 - 4 \frac{ \lambda }{ L }
	+ 12 \frac{ \lambda^2 }{ L^2 } + \mathcal{O}( \lambda/ L)^3
	\label{eq:series-scaling-function}
\end{equation}

\section{Conclusions}

We have analyzed the Casimir entropy for the ideal electron gas, in particular 
the contribution of electric currents frozen inside the bulk. This system shows
(electromagnetic) disorder, and 
the third law of thermodynamic does not apply in its orthodox formulation. We have
recovered the `entropy defect' (negative Casimir entropy at zero temperature)
reported in several places in the literature. Its thermodynamically consistent interpretation 
is the measure of the change in the disorder of the frozen currents due to their 
interaction through quasi-static magnetic fields.

\paragraph{Acknowledgments.}
We acknowledge financial support by the European Science Foundation 
within the activity `New Trends and Applications of the Casimir 
Effect' (www.casimir-network.com). F.I.\  acknowledges financial support by the Alexander von Humboldt Foundation.

\end{document}